\documentclass[showpacs,prb,aps,superscriptaddress,twocolumn]{revtex4}

\usepackage{graphicx}
\usepackage{color}

\begin{document}

\title{Shot noise in magnetic tunneling structures with two-level quantum dots}

\author{T. Szczepa\'nski}
\affiliation{Department of Physics and Medical Engineering,
Rzesz\'ow University of Technology, al.~Powsta\'nc\'ow Warszawy 6,
35-959 Rzesz\'ow, Poland}

\author{V. K. Dugaev}
\affiliation{Department of Physics and Medical Engineering,
Rzesz\'ow University of Technology, al.~Powsta\'nc\'ow Warszawy 6,
35-959 Rzesz\'ow, Poland}

\author{J. Barna\'s}
\affiliation{Department of Physics, Adam Mickiewicz University,
Umultowska~85, 61-614 Pozna\'n, Poland}
\affiliation{ Institute of Molecular Physics, Polish Academy of Sciences,
ul. M. Smoluchowskiego 17, 60-179 Pozna\'n, Poland}

\author{I. Martinez}
\affiliation{Dpto. de Fisica de la Materia Condensada, C-III, IFIMAC and INC, Universidad Autonoma de Madrid, 28049, Madrid, Spain}

\author{J. P. Cascales}
\affiliation{Dpto. de Fisica de la Materia Condensada, C-III, IFIMAC and INC, Universidad Autonoma de Madrid, 28049, Madrid, Spain}
\affiliation{Francis Bitter Magnet Lab, Massachusetts Institute of Technology, Cambridge, Massachusetts 02139, USA}

\author{J.-Y. Hong}
\affiliation{Department of Physics, National Taiwan University, 10617 Taipei, Taiwan}

\author{M.-T. Lin}
\affiliation{Department of Physics, National Taiwan University, 10617 Taipei, Taiwan}
\affiliation{Institute of Atomic and Molecular Sciences, Academia Sinica, Taipei 10617, Taiwan}

\author{F. G. Aliev}
\affiliation{Dpto. de Fisica de la Materia Condensada, C-III,  IFIMAC and INC, Universidad Autonoma de Madrid, 28049, Madrid, Spain}

\date{\today}

\begin{abstract}
We analyze shot noise in a magnetic tunnel junction with a two-level quantum dot attached to the magnetic electrodes.
The considerations are limited to the case
when some transport channels are suppressed at low temperatures. Coupling of the two dot's levels
to the electrodes are assumed to be generally different and also spin-dependent. To calculate the shot noise we
apply the approach based on the full counting statistics. The approach is used to account for  experimental
data obtained in magnetic tunnel junctions with organic barriers.
The experimentally observed  Fano factors correspond to the super-Poissonian statistics, and also depend
on relative orientation  of the electrodes' magnetic moments. We have also calculated the corresponding spin shot noise,
which is associated with fluctuations of spin current.
\end{abstract}

\maketitle


\section{Introduction}

The problem of current fluctuations has been attracting recently more and more attention
due to increasing role of fluctuations of various physical quantities in the nanoworld.\cite{kogan96,blanter00,nazarov03}
In principle, this is rather obvious because the fluctuations strongly increase with decreasing number
of particles in the system.\cite{landau5}
Starting from the pioneering article by Schottky \cite{schottky1918} and several famous papers of Khlus,\cite{khlus87}
Lesovik \cite{lesovik89} and B\"uttiker et al., \cite{buttiker90,beenakker92}
the theoretical study of current fluctuations became an exciting field of research
in statistical physics. One of the most impressive achievements of the theory
is the full counting statistics\cite{levitov92,levitov93,ivanov93,levitov96,bagrets03,levitov04,belzig05,kaasbjerg15},
which allows to calculate the
correlation functions of any order and to identify the type of statistics of current correlations.

In addition, recent progress in experimental methods has resulted in modern measurement techniques which allow to study experimentally
the current noise and extract from the noise  even more information than from
the usual measurement of the average current.\cite{lu03,fujisawa06,flindt09} Obviously, this concerns not only
the fluctuations of current, but also fluctuations of any transport-related quantity like, for example, spin or pseudospin
current, spin torque, heat fluxes, and others.

It is well known that there are various  sources of the noise. Correspondingly, the dominant
mechanism always depends on a specific problem under consideration and on various additional internal and external factors.
Here we consider the shot noise which has purely quantum character. The shot noise is mostly observed at
low-temperatures, where the corresponding experimental data show that it does not depend on temperature and is also constant
in the low-frequency range.\cite{blanter00}
The mechanism of shot noise is related to the quantization of
charge and spin of particles, that are transferred through the system.

The methods used for theoretical treatment of the noise are also different, depending on
the role of Coulomb interaction, phonons, disorder, etc. It turned out that in
some cases one can formulate a general approach which is based on the master equation
describing dynamics of quantum states of the system, so that the correlation functions (so-called cumulants)
describing current correlations in all orders (not only pair correlations) can be derived from a
single generating function.
The method of such calculations of cumulants is known as the full counting statistics\cite{levitov93,levitov96}
(FCS), and it provides a complete  description of fluctuations in
the system. In particular, one can find the mean value of current, zero-frequency
pair correlation function (shot noise), and also establish the statistics of
fluctuations -- whether it is Poissonian or any other (super-Poissonian or sub-Poissonian).
Some examples of using this method are presented in Ref.~\onlinecite{bagrets03}.

The approach based on FCS was used to  explain the super-Poissonian shot noise, for which
the Fano factor $F$ in a tunnel junction with quantum dot \cite{belzig05} is higher, $F>1$,
than the corresponding Fano factor for Poissonian statistics ($F=1$).
Here, we consider a similar problem of current and spin current noise in a magnetic tunnel junction
with a nonmagnetic quantum dot, but the dot is attached to two ferromagnetic electrodes.
The experiments on organic tunnel junctions with ferromagnetic contacts demonstrated
super-Poissonian shot noise which additionally depends on magnetic polarization of the electrodes.\cite{cascales14}
It was assumed that the model based on  transfer of electrons through
two discrete levels of molecules is sufficient to describe statistics of the fluctuations in such a
system.\cite{cascales14}

It should be also noted that the problem of spin shot noise has been already considered in
many papers~\cite{mishchenko03,belzig04,lamacraft04,wang04,sauret04,foros05,guerrero06,dragomirova07,chudnovskiy08,
cascales12,szczepanski13,tang14,xue15_1,burtzlaff15,arakawa15}
and for various systems. The main interest  of these works was focused on how the discreteness of spin affects the
current fluctuations.

In this paper we present a theoretical description of the model used for explanation of
the experimental data on magnetic tunnel junctions with organic molecules.\cite{cascales14}
Apart from charge fluctuations, we also consider spin fluctuations
which influence the electric current. Moreover, we also consider how the spin fluctuations affect the spin
current in the system. In Sec.~II we describe the model and the theoretical
method used to calculate the noise. Current shot noise is calculated in Sec.~III,
while the spin noise is calculated in Sec.~IV. The relation with the experiment is discussed in Sec. V,
and the discussion of results and final conclusions are in Sec.~VI.

\section{Model and theoretical method}

The  model considered in this paper is based on a quantum dot with two discrete electron levels~\cite{molecule}
coupled {\it via} tunneling processes to the  left and right magnetic electrodes. We assume
that the direct tunneling between the electrodes (so-called cotunneling) is very small as compared to the sequential tunneling through
the levels of the quantum dot, and therefore will be ignored.
Apart from this, Coulomb interaction of electrons localized at the dot is assumed to be strong enough to
completely suppress the states with two electrons in the dot. This model is a direct generalization
of the model studied in Ref.~\onlinecite{belzig05} to the case of a magnetic junction
-- two magnetic leads and a non-magnetic quantum dot. Accordingly, we assume (i)
different probabilities for tunneling of spin-up and spin-down electrons from the dot to the leads (and {\it vice versa}), and
(ii) different probabilities of tunneling from/to the low-energy and high-energy levels of the dot.
The system under consideration is shown schematically in Fig.~1. The central part presents the two-level system,
and both energy levels are coupled to the leads {\it via} the hopping terms. We consider the situation
when the system is biased as shown in Fig.~1, so electrons tunnel from right to left.

\begin{figure}
\includegraphics[width=0.8 \linewidth]{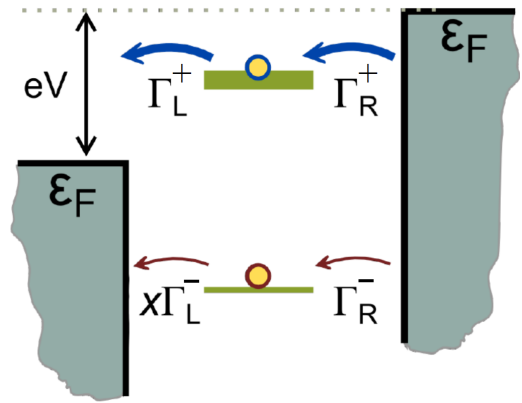}
\caption{Schematic of the tunnel junction with two-level quantum dot considered in this paper.}
\end{figure}

The key property of the model~\cite{belzig05} is an assumption that the
low-energy level $\varepsilon _-$ of the dot
is below the Fermi level of left electrode (and thus also of the right electrode), as shown schematically in Fig.~1. Hence,
at $T=0$ there is no tunneling of electron to the left (and also to the right) from the dot, and the
junction is completely blocked. At nonzero temperatures there are possible hopping processes to the left,
which should be taken
into account. This is accounted for by a temperature-dependent factor $x$ which describes
tunneling to the left at the energy, at which
all electron states in the left electrodes are filled at $T=0$ (Fig.~1), but may be empty at higher temperatures.
We consider the case of $T\ne 0$ and assume that the density
of temperature-activated holes in the left electrode is relatively small, so the parameter
$x$ can be evaluated as $x\sim \exp[(\varepsilon _{-}-E_F)/k_BT]\ll 1$.

To calculate the shot noise in junctions under consideration, we follow the method of FCS calculations proposed
by  Bagrets and Nazarov.\cite{bagrets03} First, we need to find the probability of quantum dot to be in one of the possible
quantum states, which can be
found from the following master equation describing dynamics of the dot's states:
\begin{eqnarray}
\label{1}
\frac{d{\bf P}}{dt}=\hat{M}{\bf P},
\end{eqnarray}
where
\begin{eqnarray}
\label{2}
{\bf P}^T=\left( P^-_\uparrow ,\, P^-_\downarrow ,\, P^+_\uparrow ,\, P^+_\downarrow ,\, P_0 \right)
\end{eqnarray}
is a vector whose components describe probabilities of the dot to be in the state with one
spin-$\sigma $ electron in
the low-energy level ($P^-_\sigma $), one spin-$\sigma $ electron in
the high-energy level ($P^+_\sigma $), and the probability of the state with no electrons in the dot ($P_0$).

As already mentioned above, the state with two excess electrons in the dot is assumed to have rather high energy due to strong electron
correlations, so it is ruled out from the considerations. This assumption is well justified when QDs are sufficiently small.
In our case we consider tunneling through short molecules which play a role of QDs. Coulomb energy of doubly charged molecules is then sufficiently large, so the above assumption is reasonable and well justified.

The matrix $\hat{M}$ on the right side of the master equation (1) includes the
rates $\Gamma ^\pm_{L\sigma }$ and
$\Gamma ^\pm_{R\sigma }$ of electron tunneling from the dot to the left electrode and
from the right electrode to the dot, respectively,
\begin{eqnarray}
\label{3}
\hat{M}=\left( \begin{array}{ccccc}
-x\Gamma ^-_{L\uparrow} & 0 & 0 & 0 & \Gamma ^-_{R\uparrow} \\
0 & -x\Gamma ^-_{L\downarrow} & 0 & 0 & \Gamma _{R\downarrow}^- \\
0 & 0 & -\Gamma ^+_{L\uparrow} & 0 & \Gamma ^+_{R\uparrow} \\
0 & 0 & 0 & -\Gamma ^+_{L\downarrow} & \Gamma ^+_{R\downarrow} \\
x\Gamma ^-_{L\uparrow} & x\Gamma ^-_{L\downarrow} & \Gamma ^+_{L\uparrow} &
\Gamma ^+_{L\downarrow} & -\Gamma _\Sigma
\end{array} \right) ,
\end{eqnarray}
where we also introduced the notation
$\Gamma _\Sigma =\Gamma _{R\uparrow}^++\Gamma _{R\uparrow}^-+\Gamma _{R\downarrow}^++\Gamma _{R\downarrow}^-$.
Since the electrodes are ferromagnetic, the tunneling probabilities are assumed to be dependent
on the electron spin orientation.
The signs ascribed to the elements of the matrix $\hat{M}$ correspond to increasing
or decreasing probability of the corresponding dot state due to the respective tunneling processes.
The factor $x$ in this matrix was already defined above and is assumed to be small,
$x\ll 1$.

To distinguish between the probabilities of electron tunneling from the right electrode to the
upper or to the lower energy level of the dot, we introduced different  parameters $\Gamma _{R\sigma }^+$
and $\Gamma _{R\sigma }^-$. This difference can be attributed to different shapes of the
electronic orbitals corresponding to the dot's states.
Transmission of electrons in the tunneling structure shown in Fig.~1 is a stochastic process, which consists of random
hoppings of electrons between electrodes and QD at random
times $\tau _i$. Therefore, the calculation of mean current, say through the left junction, as well as
of current correlation functions imply averaging over processes $\zeta _s$ with an arbitrary number $s$ of sequential transitions with electron
transfer in all possible channels. The probability $Q_s$ of the process $\zeta _s$ is determined by the
probabilities of system to stay in certain states during the time between transitions and by the probability
of single transitions at  $\tau _i$ ($i=1,...s$) specified by the process $\zeta _s$. The probabilities of particular
transitions are the matrix elements in Eq.~(3). To find the generating function $S(\chi )$ of cumulant expansion
one has to average the exponent $\exp \{ i\int d\tau \, \chi (\tau )\, \hat{I}(\tau )\} $, where $\hat{I}(\tau )$ is an
instanteneous current at $\tau $ and $\chi (\tau )$ is the source field introduced to find
current cumulants by using the generating function $S(\chi )$. The key point of the theory in Ref.~\onlinecite{bagrets03} is
that averaging of expression for the generating function with source field $\chi $ induces
$\chi $-dependent probabilities $Q^\chi _s$ which differ from $Q_s$ by an exponential factor $e^{i\chi (\tau )}$ in
the probability of tunneling through the considered (left) junction. All the details of this derivation can be found in the cited work.

Thus, following the method of Ref.~\onlinecite{bagrets03}, we consider eigenvalues of the matrix
$\hat{Z}(\chi )$ defined as
\begin{eqnarray}
\label{4}
\hat{Z}(\chi )=\left( \begin{array}{ccccc}
-x\Gamma ^-_{L\uparrow} & 0 & 0 & 0 & \Gamma _{R\uparrow}^- \\
0 & -x\Gamma ^-_{L\downarrow} & 0 & 0 & \Gamma _{R\downarrow}^- \\
0 & 0 & -\Gamma ^+_{L\uparrow} & 0 & \Gamma ^+_{R\uparrow} \\
0 & 0 & 0 & -\Gamma ^+_{L\downarrow} & \Gamma ^+_{R\downarrow} \\
x\Gamma ^-_{L\uparrow}e^{i\chi } & x\Gamma ^-_{L\downarrow}e^{i\chi } & \Gamma ^+_{L\uparrow}e^{i\chi } &
\Gamma ^+_{L\downarrow}e^{i\chi } & -\Gamma _\Sigma
\end{array} \right) .\hskip0.4cm
\end{eqnarray}
As compared to $\hat{M}$, the matrix $\hat{Z}(\chi )$ includes an additional phase factor $e^{i\chi }$,
which allows to determine the generating function $S(\chi )$ of the current correlators,
\begin{eqnarray}
\label{5}
S(\chi )=-t_0\lambda _{0}(\chi ),
\end{eqnarray}
where $t_0$ is the period of transfer of a charge, and
$\lambda _0(\chi )$ is the lowest eigenvalue of the matrix $\hat{Z}(\chi )$,
\begin{eqnarray}
\label{6}
\det [\hat{Z}(\chi )-\lambda ] =0.
\end{eqnarray}

In the case of $x=0$ (which corresponds to $T=0$) one obtains from Eq.~(6) that the minimum eigenvalue of $\hat{Z}(\chi )$ is
$\lambda _0=0$.
Thus, for small $x$,  $x\to 0$, one may look for a solution which is linear in $x$, $\lambda _0=x\tilde{\lambda }$.
Using then Eqs.~(4) and (6) we find the following algebraic equation for $\tilde{\lambda }$:
\begin{eqnarray}
\label{7}
(\Gamma ^-_{L\uparrow}+\tilde{\lambda })(\Gamma ^-_{L\downarrow}+\tilde{\lambda })
[(\Gamma ^+_{R\uparrow}+\Gamma ^+_{R\downarrow})(-1+e^{i\chi })-\Gamma _{R\uparrow}^--\Gamma _{R\downarrow}^-]
\nonumber \\
+e^{i\chi }(\Gamma ^-_{L\uparrow}+\tilde{\lambda })\Gamma ^-_{L\downarrow}\Gamma ^-_{R\downarrow}
+e^{i\chi }(\Gamma ^-_{L\downarrow}+\tilde{\lambda })\Gamma ^-_{L\uparrow}\Gamma ^-_{R\uparrow}=0. \hskip0.3cm
\end{eqnarray}
This is a quadratic equation for $\tilde{\lambda }$, which can be presented as $\tilde{\lambda }^2+2b\tilde{\lambda }+c=0$, where
\begin{eqnarray}
\label{8}
b&=& \frac{(\Gamma ^-_{L\uparrow}\Gamma ^-_{R\uparrow}+\Gamma ^-_{L\downarrow}\Gamma ^-_{R\downarrow})\, e^{i\chi }}
{2[(\Gamma ^+_{R\uparrow}+\Gamma ^+_{R\downarrow})(e^{i\chi }-1)-\Gamma ^-_{R\uparrow}-\Gamma ^-_{R\downarrow}]}
\nonumber \\
&&+\frac{\Gamma ^-_{L\uparrow}+\Gamma ^-_{L\downarrow}}{2}\, , \\
c&=&\Gamma ^-_{L\uparrow}\Gamma ^-_{L\downarrow}\,
\frac{(e^{i\chi }-1)(\Gamma ^+_{R\uparrow}+\Gamma ^+_{R\downarrow}+\Gamma _{R\uparrow}^-+\Gamma _{R\downarrow}^-)}
{(\Gamma ^+_{R\uparrow}+\Gamma ^+_{R\downarrow})(e^{i\chi }-1)-\Gamma ^-_{R\uparrow}-\Gamma ^-_{R\downarrow}}\, .
\end{eqnarray}
Thus, the FCS generating function in the limit of low $T$ (small $x$) can be written as
\begin{eqnarray}
\label{10}
S(\chi )=-t_0x\, \big( -b\pm \sqrt{b^2-c}\big),
\end{eqnarray}
with the parameters $b$ and $c$ defined by Eqs.~(8) and (9).

\section{Electric current shot noise}

The mean value of electric current, $\overline{I}$, and the correlator of current fluctuations (shot noise), $S_2$,
are determined by the first two cumulants $C_n$ ($n=1,2$) of the generating function,
$C_n=-(-i)^n [d^nS(\chi)/d\chi^n]|_{\chi =0}$, i.e. explicitly
\begin{eqnarray}
\label{11}
&&\overline{I}=ieS'(\chi )|_{\chi =0}
\\
&&S_2=\overline{(I-\overline{I})^2}=2e^2S''(\chi )|_{\chi =0},
\end{eqnarray}
respectively, where $S'(\chi )$ and $S''(\chi )$ stand for the first and second derivative of $S$ with respect to $\chi$.
Obviously, the FCS method gives the possibility to calculate all higher current correlation functions,
$S_3$, $S_4$, etc.

Using Eqs.~(8)-(12) one finds the following expression for the mean value of electric current (we take the units with $t_0=1$):
\begin{eqnarray}
\label{13}
\overline{I}=\frac{ex\Gamma ^-_{L\uparrow}\Gamma ^-_{L\downarrow}
(\Gamma ^+_{R\uparrow}+\Gamma ^+_{R\downarrow}+\Gamma _{R\uparrow}^-+\Gamma _{R\downarrow}^-)}
{\Gamma ^-_{L\uparrow}\Gamma _{R\downarrow}^-+\Gamma ^-_{L\downarrow}\Gamma _{R\uparrow}^-}\,. \hskip0.3cm
\end{eqnarray}
We recall that the above expression is valid in  the low temperature limit, where $x\ll 1$.
Similarly, once can also determine the relevant shot noise $S_2$. Since the corresponding formula is relatively long,
we present it in Appendix, see Eq.(A12), where we also give more details on its derivation. Having found
the shot noise, one can determine the corresponding Fano factor,
\begin{eqnarray}
\label{14}
F=\frac{C_2}{C_1}=\frac{S_2}{2e\overline{I}}
=\frac{2(\Gamma ^+_{R\uparrow }+\Gamma ^+_{R\downarrow })+\Gamma _{R\uparrow }^-+\Gamma _{R\downarrow }^-}
{(\Gamma _{R\uparrow }^-+\Gamma _{R\downarrow }^-)}\hskip0.5cm
\nonumber \\
+\frac{2(\Gamma ^-_{L\uparrow }\Gamma _{R\uparrow }^-+\Gamma ^-_{L\downarrow }\Gamma _{R\downarrow }^-)
(\Gamma ^+_{R\uparrow }+\Gamma ^+_{R\downarrow }+\Gamma _{R\uparrow }^-+\Gamma _{R\downarrow }^-)}
{(\Gamma _{R\uparrow }^-+\Gamma _{R\downarrow }^-)
(\Gamma ^-_{L\uparrow }\Gamma _{R\downarrow }^-+\Gamma ^-_{L\downarrow }\Gamma _{R\uparrow }^-)}
\nonumber \\
-\frac{2\Gamma ^-_{L\uparrow }\Gamma ^-_{L\downarrow }
(\Gamma ^+_{R\uparrow }+\Gamma ^+_{R\downarrow }+\Gamma _{R\uparrow }^-+\Gamma _{R\downarrow }^-)
(\Gamma _{R\uparrow }^-+\Gamma _{R\downarrow }^-)}
{(\Gamma ^-_{L\uparrow }\Gamma _{R\downarrow }^-+\Gamma ^-_{L\downarrow }\Gamma _{R\uparrow }^-)^2}\, .
\end{eqnarray}

In the nonmagnetic case, $\Gamma _{L\uparrow }^\pm =\Gamma _{L\downarrow }^\pm =\Gamma _{L}$
and $\Gamma _{R\uparrow }^\pm =\Gamma _{R\downarrow }^\pm =\Gamma _{R}$, we obtain the results
of Ref.~\onlinecite{belzig05}, with the lowest two cumulants and the Fano factor equal
\begin{eqnarray}
\label{15}
C_1=2x\Gamma _{L},\, C_2=6x\Gamma _{L},\, {\rm and}\,\; F=C_2/C_1=3.
\end{eqnarray}
Thus, the corresponding shot noise is then super-Poissonian, with $F=3$.
If we take into account the spin dependence of electron tunneling, but assume
$\Gamma _{R\sigma }^-=\Gamma _{R\sigma }^+$, then we find from Eq.~(15) that the Fano factor
is even larger than 3, $F>3$, for any choice of other parameters.

\begin{figure}
\includegraphics[width=1.0 \linewidth]{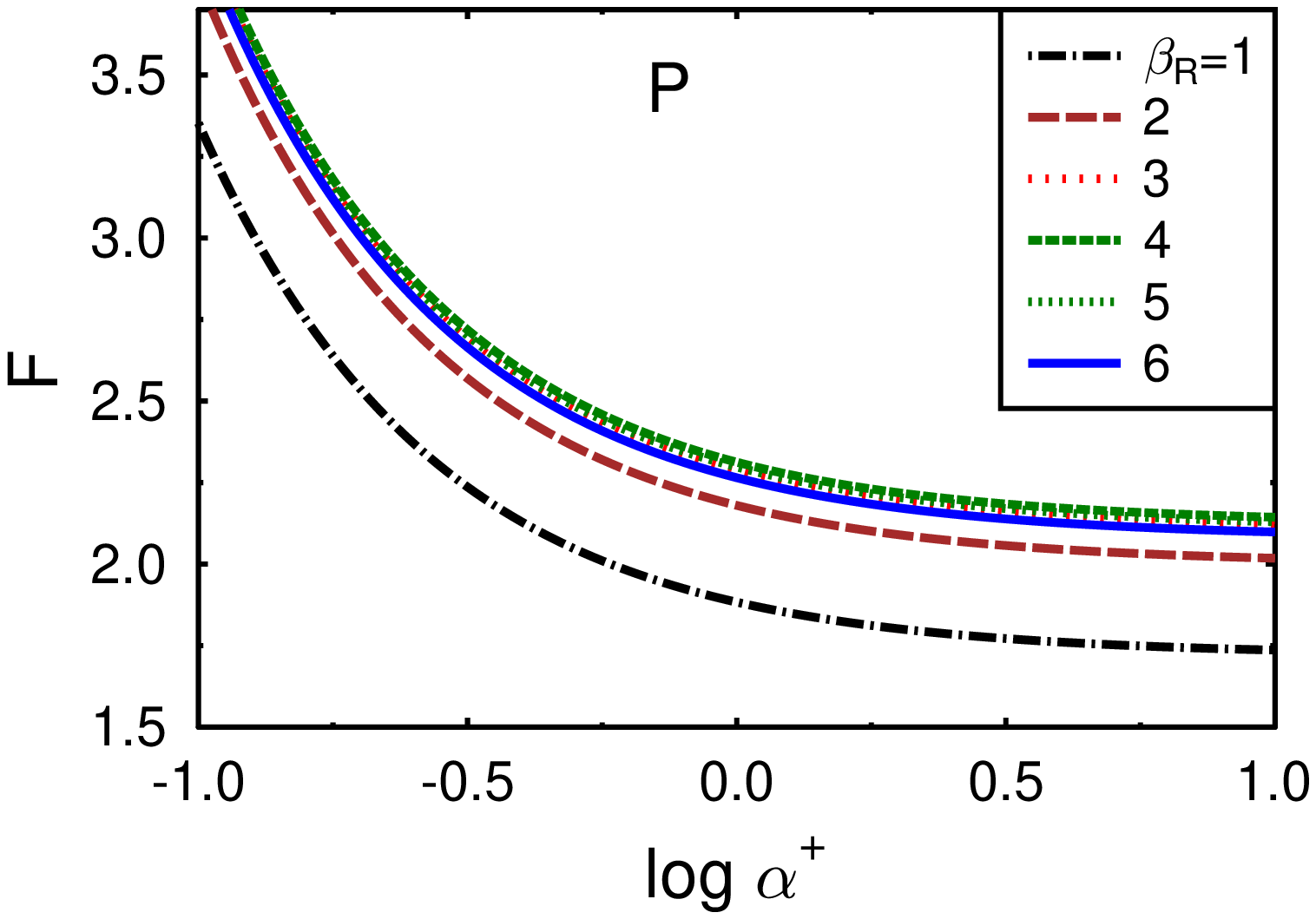}
\includegraphics[width=1.0 \linewidth]{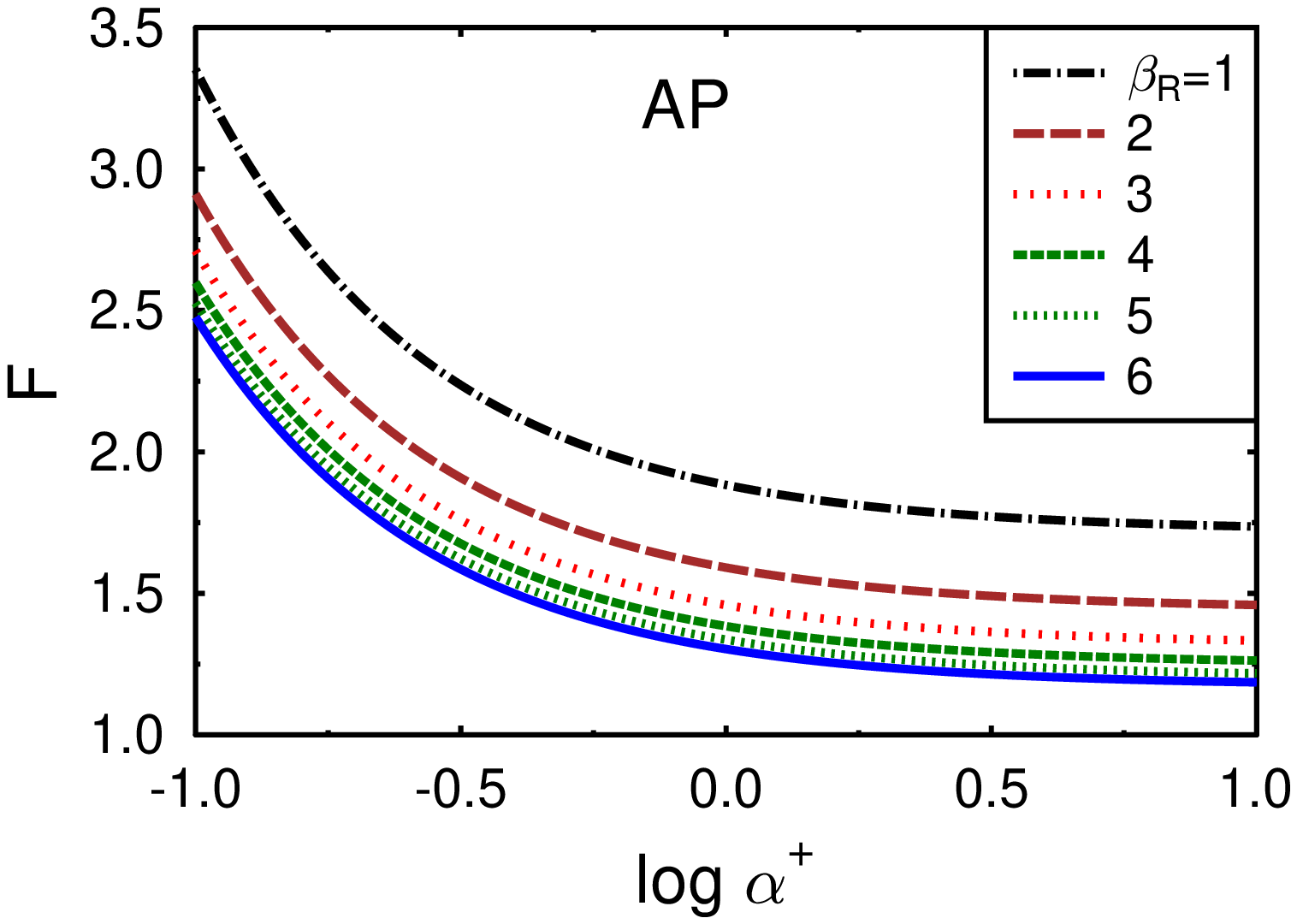}
\caption{Fano factor in the parallel (top) and antiparallel (bottom) magnetic configurations
as a function of $\alpha ^+$.
The other parameters are $x_R=0.3$, $x_L=1$, $\beta _L=4$, $\alpha ^-=0.2$, and $\beta _R$ as indicated.}
\end{figure}

One can describe the shot noise and the Fano factor (15) by a certain number of parameters which
quantify the relevant asymmetry in each of the transport channels. To do this let us define the junction
resistance  $R_{{L,R}\sigma }^\pm $ for each level- and spin-channel.  The resistance $R_{{L,R}\sigma }^\pm $ is inversely
proportional to the corresponding tunneling rate $\Gamma _{{L,R}\sigma }^\pm $.
Accordingly, we introduce the parameters $\alpha ^+=R_{R\uparrow }^+/R_{L\uparrow }^+$ and
$\alpha ^-=R_{R\uparrow }^-/R_{L\uparrow }^-$ to describe the right-left asymmetry, in the spin-up channel
associated with  the high-energy and low-energy dot's levels. Apart from this, we also define the
parameters  $\beta _R=R_{R\downarrow }^-/R_{R\uparrow }^-$ and $\beta _L=R_{L\downarrow }^-/R_{L\uparrow }^-$
for the spin asymmetry in the coupling of the low-energy dot's level to the leads. To describe  a difference
in the coupling of the two levels of the dot to the right electrode, we introduce the parameter $x_R$ defined
as $x_R=R_{R\uparrow}^-/R_{R\uparrow}^+$. Similar parameter is also introduced to describe asymmetry of the
coupling of the two levels to the left electrode, $x_L=R_{L\uparrow}^-/R_{L\uparrow}^+$.

In case of magnetic electrodes, we also distinguish between the parallel (P)
and antiparallel (AP)
arrangements of the magnetic moments of both electrodes.
For definiteness, we define the spin-up orientation as the
orientation of majority spins  in the left electrode (i.e., opposite to magnetization vector in the
left electrode), and assume that magnetic moment of the right electrode is reversed in the AP configuration.
Thus, in the AP configuration the spin-up and spin-down electrons in the
right electrode correspond to the spin-minority and spin-majority electrons, respectively.

\begin{figure}
\includegraphics[width=1.0 \linewidth]{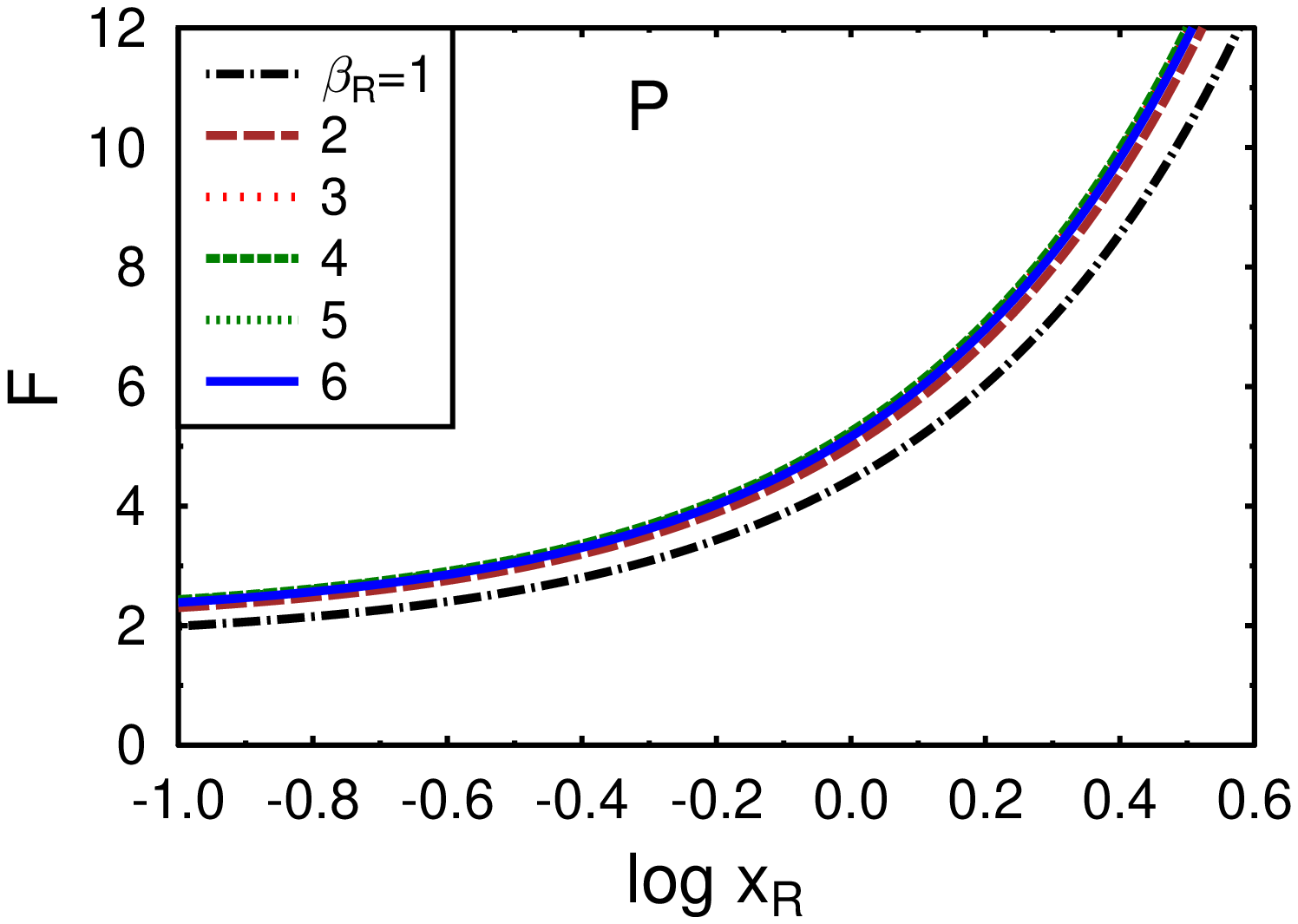}
\includegraphics[width=1.0 \linewidth]{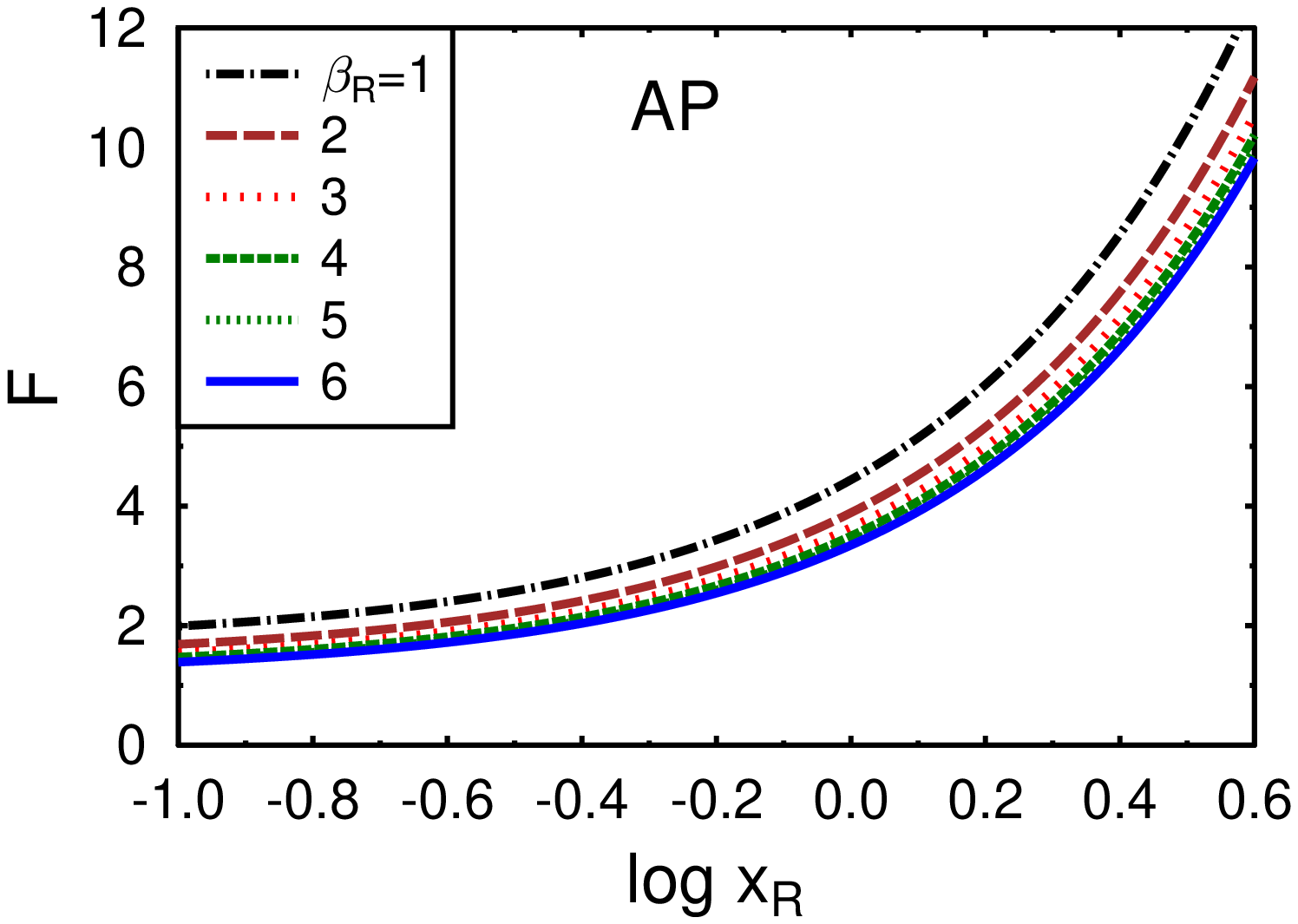}
\caption{Fano factor in the parallel (top) and antiparallel (bottom) configurations as a function
of $x_R$
for $\alpha ^+=\alpha ^-=1$, $x_L=1$, $\beta _L=4$, and $\beta _R$ as indicated.}
\end{figure}

Variation of the Fano factor $F$ with the parameter $\alpha ^+=R_{R\uparrow }^+/R_{L\uparrow }^+$
in the P and AP configurations is shown in Fig.~2 for different
values of the parameter $\beta _R$. Two features immediately follow from this figure. First, the shot noise and
thus also the Fano factor are strongly enhanced
when $\alpha^+<<1$, i.e. for $R_{R\uparrow }^+<<R_{L\uparrow }^+$. This is because a spin-up electron
tunneling from the source (right) electrode to the high-energy level spends relatively long time before tunneling
further to the sink (left) electrode, blocking this way electronic transport {\it via} other channels.
Second, the Fano factor in the parallel configuration is generally larger than in the antiparallel state.
Note, that for $\beta_R=1$ the parallel and antiparallel configurations are equivalent (right electrode
is then nonmagnetic). Then, when $\beta_R>1$, the Fano factor in the parallel configuration is lower
while in the antiparallel state is higher, which is in agreement with earlier observations.\cite{cascales14}
In turn, dependence of the Fano factor on the parameters  $x_R=R_{R\uparrow}^-/R_{R\uparrow}^+$
is shown in Fig.~3 for both magnetic configurations. The noise is super-Poissonian and the Fano factor
is relatively large for  $x_R>>1$, i.e. for $R_{R\uparrow}^->>R_{R\uparrow}^+$. Again, the noise is
smaller in the antiparallel configuration.

\begin{figure}
\includegraphics[width=1.0 \linewidth]{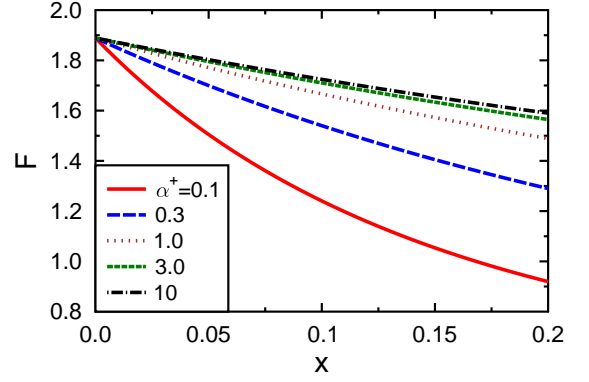}
\caption{Dependence of the Fano factor on the temperature factor $x$ in the P configuration for
$\alpha ^{-}=0.3$ and different values of $\alpha ^{+}$.
The other parameters are  $x_R=0.3$, $x_L=1$, $\beta _R=2$, and $\beta _L=4$.}
\end{figure}

When the temperature increases,  the parameter $x$ in Eq.~(3) also increases, which leads to
the temperature dependence of the Fano factor. The simple algebraic method presented above can not be used now.
Hence, we calculated numerically the eigenvalues of the matrix $\hat{Z}(\chi )$, Eq.~(4), and used the lowest eigenvalue $\lambda _0$
of the matrix $\hat{Z}(\chi )$ to determine the first two cumulants and thus the Fano factor, $F=C_2/C_1$.
The dependence of $F$ on the temperature-dependent  parameter $x$ is presented in Fig.~4.
The low-temperature limit of the Fano factor  $F$ corresponds to  $x\to 0$. The magnitude of Fano factor essentially decreases
with increasing temperature. This is related to de-blocking of the conduction channel through the low-energy level $E_-$.
Note, the system may go to the sub-Poissonian regime with increasing temperature.

\section{Spin current noise}

The FCS method for calculation of current and current noise can be easily generalized to study
the spin current and spin current noise.
To do this  we consider the eigenvalues of the matrix
$\hat{Z}_s(\chi )$, which we define as
\begin{eqnarray}
\label{16}
\hat{Z}_s(\chi )=\left( \begin{array}{ccccc}
-x\Gamma ^-_{L\uparrow} & 0 & 0 & 0 & \Gamma _{R\uparrow}^- \\
0 & -x\Gamma ^-_{L\downarrow} & 0 & 0 & \Gamma _{R\downarrow}^- \\
0 & 0 & -\Gamma ^+_{L\uparrow} & 0 & \Gamma ^+_{R\uparrow} \\
0 & 0 & 0 & -\Gamma ^+_{L\downarrow} & \Gamma ^+_{R\downarrow} \\
x\Gamma ^-_{L\uparrow}e^{i\chi } & x\Gamma ^-_{L\downarrow}e^{-i\chi } & \Gamma ^+_{L\uparrow}e^{i\chi } &
\Gamma ^+_{L\downarrow}e^{-i\chi } & -\Gamma _\Sigma
\end{array} \right) .\hskip0.2cm
\end{eqnarray}
In contrast to Eq.~(4), we count here the hopping through
the left junction  of spin-up and spin-down electrons, corresponding to the  plus and minus sings in the exponents in the
bottom row, respectively. This  means that we calculate
the spin current as a difference of the fluxes of electrons in the spin-up and spin-down channels.

\begin{figure}
\includegraphics[width=1.0 \linewidth]{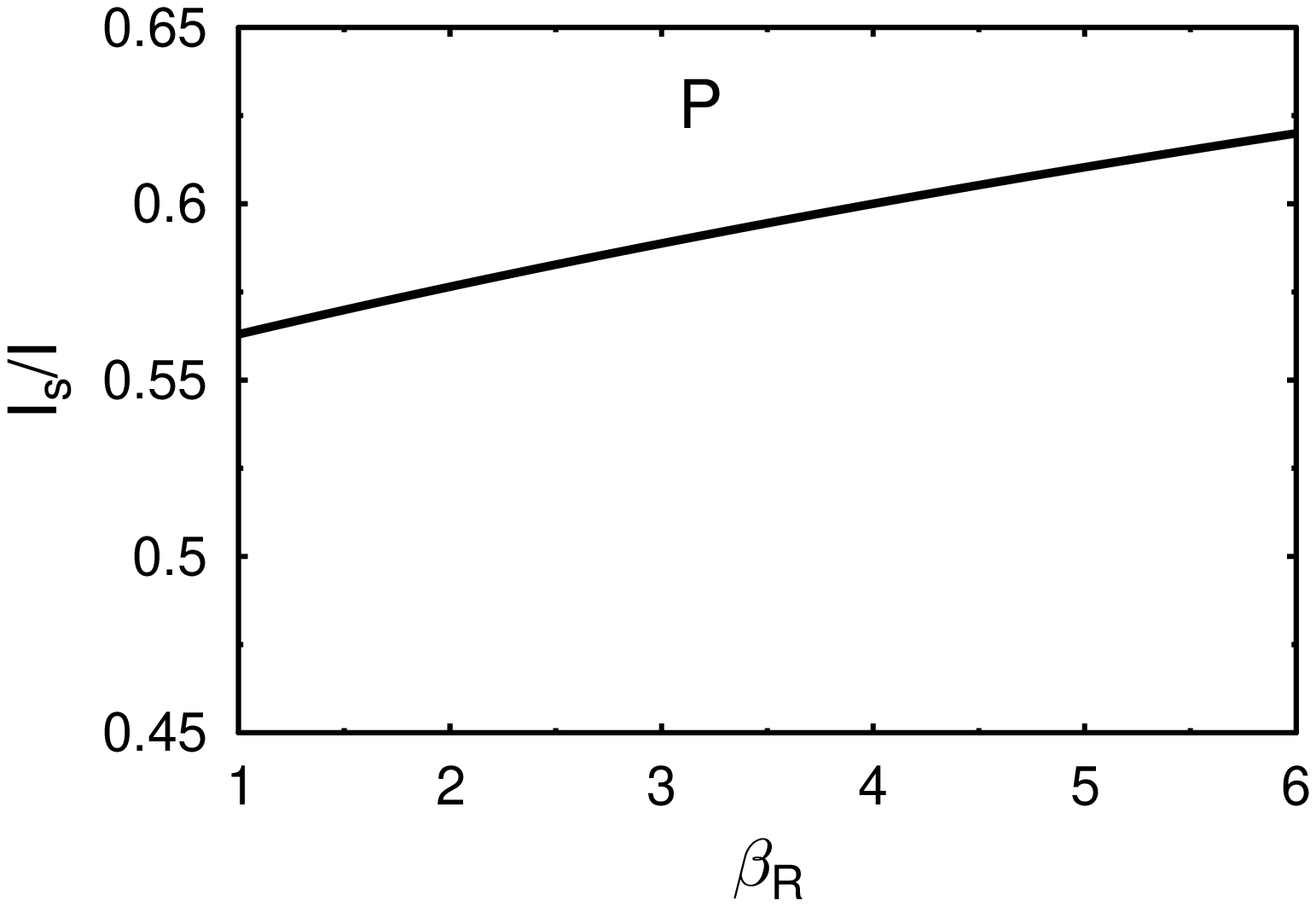}
\includegraphics[width=1.0 \linewidth]{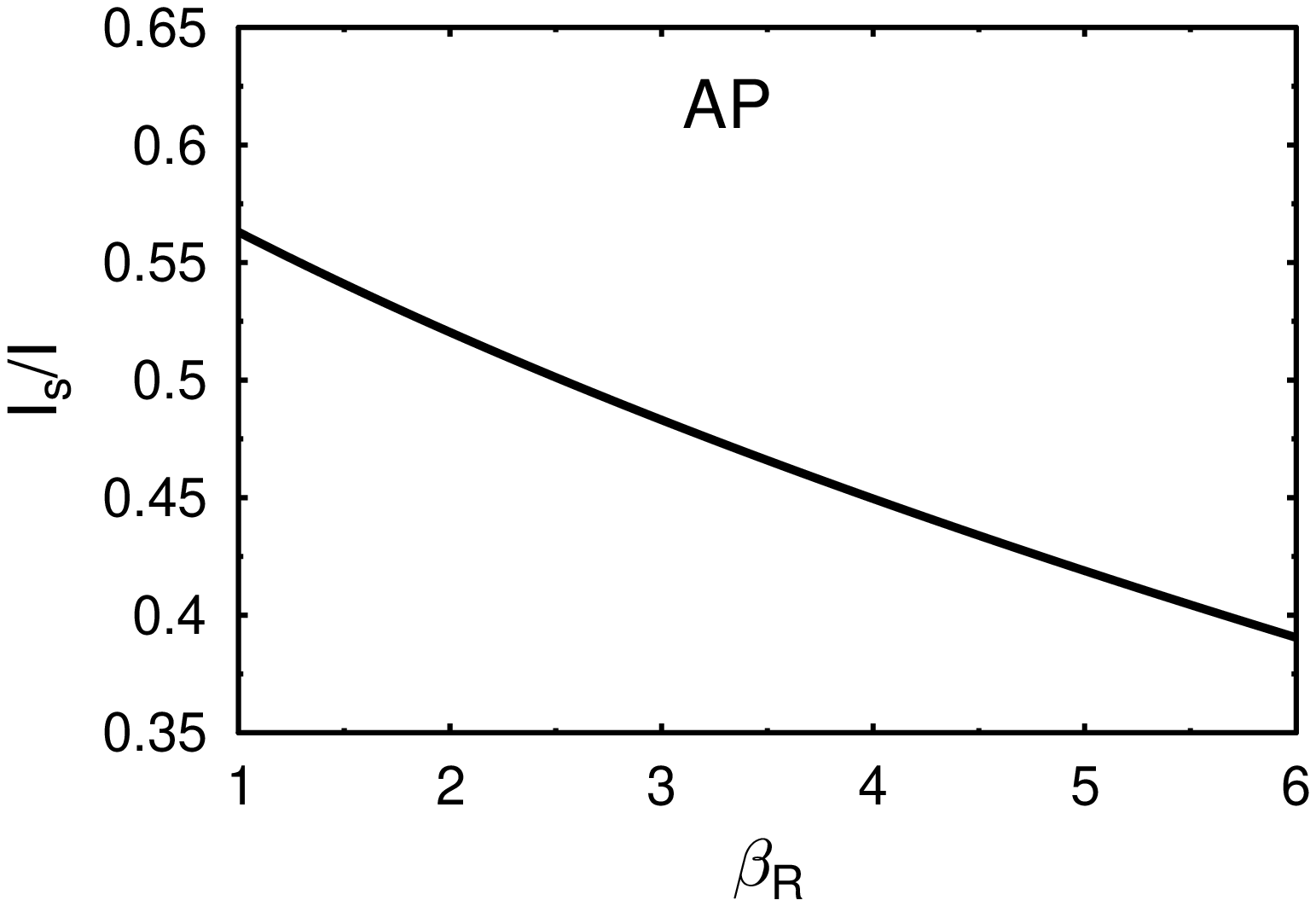}
\caption{Spin polarization of current $I_s/I$ as a function for $\beta _R$ in P (top) and AP (bottom) configurations.
The other parameters are $x_R = 0.3$, $x_L=1$, $\alpha^+ = 0.3$, $\alpha^{-}=0.2$, and $\beta_L=4$.}
\end{figure}

All the calculations are similar to those in the case of electric current, so we will not
repeat the details, but present only some results.
In the low-temperature limit ($x\to 0$) the first
two cumulants can be written in the form
\begin{eqnarray}
\label{17}
C_1^s\simeq \frac{x\Gamma ^-_{L\uparrow }\Gamma ^-_{L\downarrow }
(\Gamma ^+_{R\uparrow }+\Gamma ^-_{R\uparrow }-\Gamma ^+_{R\downarrow }-\Gamma ^-_{R\downarrow })}
{\Gamma ^-_{L\uparrow }\Gamma ^-_{R\downarrow }+\Gamma ^-_{L\downarrow }\Gamma ^-_{R\uparrow }}\, ,
\hskip0.5cm
\\
C_2^s\simeq \frac{2x\Gamma ^-_{L\uparrow }\Gamma ^-_{L\downarrow }}
{\Gamma ^-_{L\uparrow }\Gamma ^-_{R\downarrow }+\Gamma ^-_{L\downarrow }\Gamma ^-_{R\uparrow }}
\left\{ \frac12 (\Gamma ^+_{R\downarrow }+\Gamma ^-_{R\downarrow }+\Gamma ^+_{R\uparrow }+\Gamma ^-_{R\uparrow })
\right. \nonumber \\ \left.
+(\Gamma ^+_{R\downarrow }+\Gamma ^-_{R\downarrow }-\Gamma ^+_{R\uparrow }-\Gamma ^-_{R\uparrow })
\frac{\Gamma ^+_{R\downarrow }-\Gamma ^+_{R\uparrow }}
{\Gamma ^-_{R\uparrow }+\Gamma ^-_{R\downarrow }}
\right. \nonumber \\ \left.
+\left[
\Gamma ^-_{L\downarrow }\Gamma ^-_{R\downarrow }-\Gamma ^-_{L\uparrow }\Gamma ^-_{R\uparrow }
+(\Gamma ^-_{L\downarrow }\Gamma ^-_{R\downarrow }+\Gamma ^-_{L\uparrow }\Gamma ^-_{R\uparrow })
\frac{\Gamma ^+_{R\downarrow }-\Gamma ^+_{R\uparrow }}
{\Gamma ^-_{R\uparrow }+\Gamma ^-_{R\downarrow }}\right]
\right. \nonumber \\ \left.
\times \frac{\Gamma ^+_{R\downarrow }+\Gamma ^-_{R\downarrow }-\Gamma ^+_{R\uparrow }-\Gamma ^-_{R\uparrow }}
{\Gamma ^-_{L\uparrow }\Gamma ^-_{R\downarrow }+\Gamma ^-_{L\downarrow }\Gamma ^-_{R\uparrow }}
-\Gamma ^-_{L\uparrow }\Gamma ^-_{L\downarrow }(\Gamma ^-_{R\uparrow }+\Gamma ^-_{R\downarrow })
\right. \nonumber \\ \left.
\times \left( \frac{\Gamma ^+_{R\downarrow }+\Gamma ^-_{R\downarrow }-\Gamma ^+_{R\uparrow }-\Gamma ^-_{R\uparrow }}
{\Gamma ^-_{L\uparrow }\Gamma ^-_{R\downarrow }+\Gamma ^-_{L\downarrow }\Gamma ^-_{R\uparrow }}\right) ^2
\right\} . \hskip0.4cm
\end{eqnarray}

\begin{figure}[b]
\includegraphics[width=1.0 \linewidth]{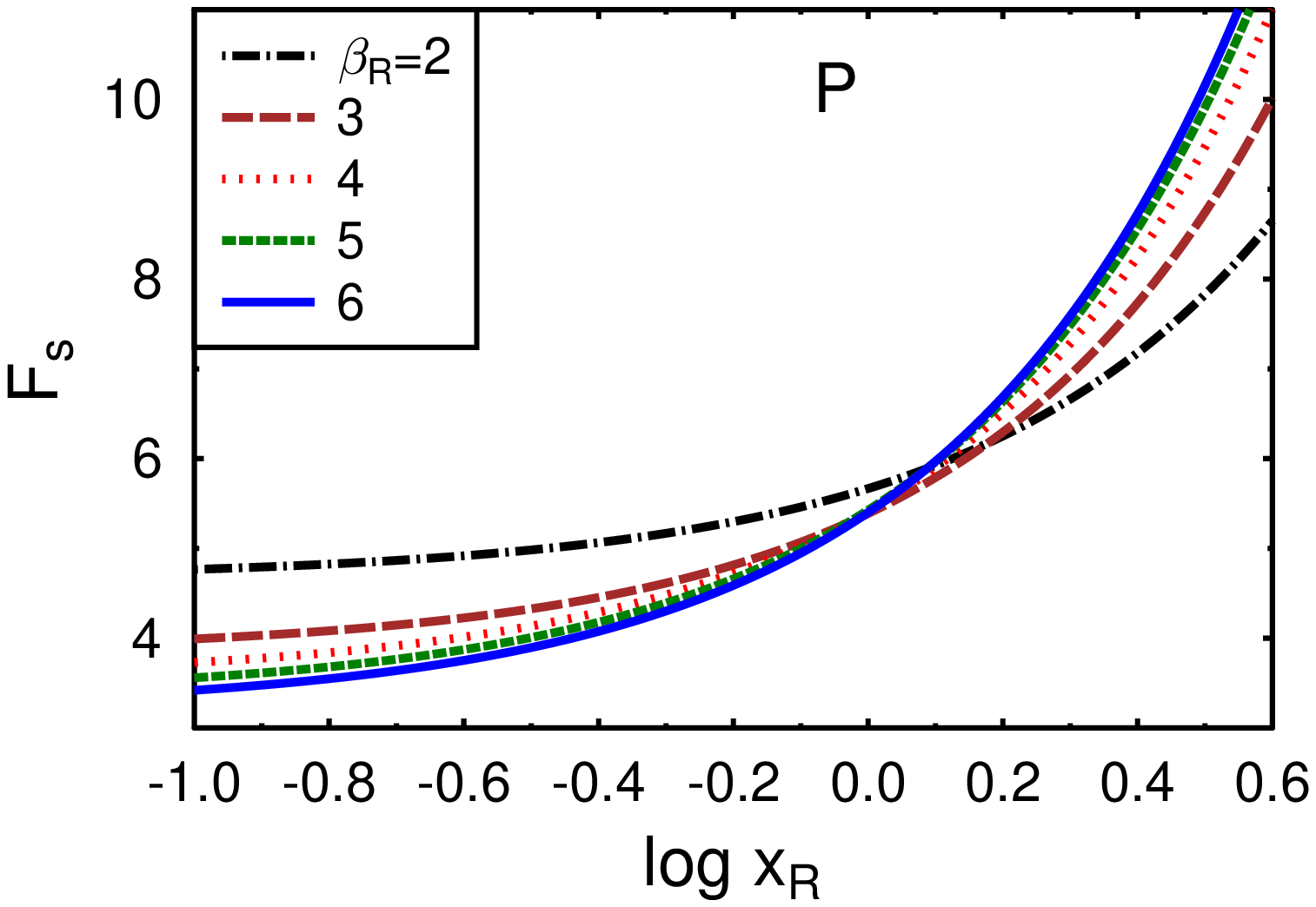}
\includegraphics[width=1.0 \linewidth]{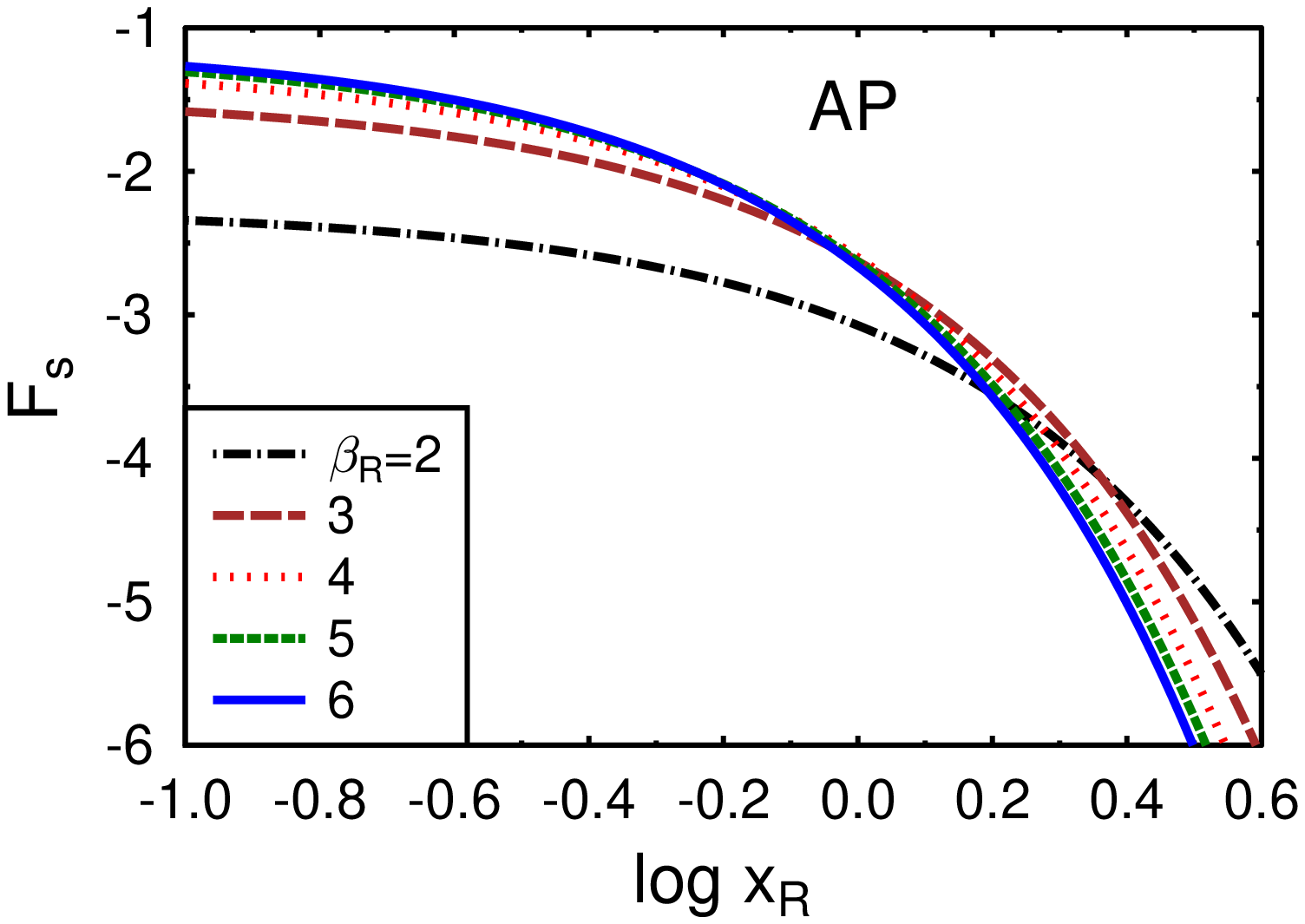}
\caption{Spin Fano factor in the parallel (top) and antiparallel (bottom) configurations as a function
of $x_R$ for $\alpha ^+=\alpha ^-=1$, $x_L=1$, $\beta _L=4$, and $\beta _R$ as indicated.}
\end{figure}

Accordingly, the mean spin current can be calculated as  $I_s=C_1^s$,
while the spin current noise as $S_2^s=2C_2^s$. In Fig.~5 we present the spin polarization
of electric current $I_s/I=C_1^s/C_1$ in the P and AP configurations. As we see, the polarization strongly depends on
the parameter $\beta _R$ describing asymmetry between the spin-up and spin-down channels.

In Fig.~6 we show the calculated  spin Fano factor, defined as $F_s=C_2^s/C_1^s$, for both
parallel and antiparallel magnetic configurations.
These Fano factors are presented as a function of  $x_R$.
In the parallel configuration the Fano factor increases with increasing $x_R$ while in the antiparallel
state it decreases with increasing $x_R$. Note, the spin Fano factor is positive in the P configuration and negative in the
AP state. This difference is associated with different signs of the spin current in the two configurations.

\section{Experimental data on shot noise in magnetic tunnel junctions}

Experimental measurements of shot noise have been performed in magnetic tunnel junctions with molecular perylene-teracarboxylic
dianhydride (PTCDA) organic barriers. The molecular layer was up to 5 nm thick.
The shot noise measurements have been done at 0.3~K and for the  bias up to 10~mV.
Detailed  description of the preparation method of the tunnel junctions and of  the experimental
technique used to measure shot noise have been published elsewhere. \cite{li11}

\begin{figure}[h]
\vspace*{0.5cm}
\includegraphics[width=0.75 \linewidth]{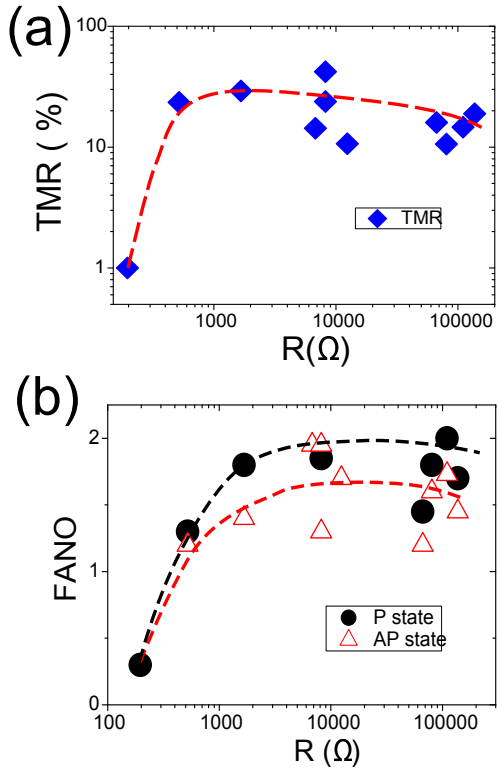}
\caption{(a) Tunneling magnetoresistance of OMTJs with PTCDA barriers with different PTCDA thickness ranging from 0~nm
(1.2~nm of AlO$_x$ buffer layers only) to 5~nm and plotted as a function of device resistance (low
bias junction resistance in the P state). Measurements have been done at
10~K and with applied bias of 1~mV  (b) Fano factor in the P and AP states as a function of OMTJs resistance measured at T=0.3~K
and averaged for the bias range about 3-10~mV. Dashed lines are guides for the eye.}
\end{figure}

Representative experimental results are shown in Fig.~7. More experimental data can be found in Ref.~\onlinecite{cascales14}.
We have measured not only the shot noise and the corresponding Fano factor, but also the tunneling magnetoresistance (TMR).
The latter is defined as the relative difference in the junction resistances in antiparallel and parallel
magnetic configurations. As one can see in Fig.~7a, the organic magnetic tunnel junctions (OMTJs) show TMR
ratio ranging between
10\% and 40\%, with the lowest value of TMR observed in the PTCDA-free samples, i.e. in the sample with no PTCDA layer,
but with  1.2~nm AlO$_x$ tunnel barrier only. The experimental values of TMR are in agreement with the model calculations
for the parameter $\beta _R\simeq 1.6$.\cite{cascales14} Note, the TMR ratio in Fig.~7 is shown as a function of low
bias junction resistance in the P state. Previous measurements indicated approximately exponential dependence of the
junction resistance on the PTCDA thickness. \cite{cascales14}

The measurements of shot noise reveal super-Poissonian tunneling statistics, with the Fano factor ranging
between 1.5 and 2 when the barrier includes the PTDCA layer (see also the preliminary report\cite{cascales14}).
The control sample (i.e., the sample without PTCDA but with 1.2~nm AlO$_x$ tunnel barrier only) shows the lowest
resistance and also the lowest Fano
factor of the order of $F$=0.3 (which corresponds to the sub-Poissonian statistics), as expected for disordered metals.
Hence, we conclude that the super-Poissonian  shot noise observed in OMTJs  is most likely associated with tunneling
through discrete states. The measured Fano factors in both magnetic configurations are shown in Fig.~7b for 3-10~mV
biased junctions. The data are also presented as a function of the junction resistance. As already reported earlier,
the  Fano factors in the AP state are smaller than in the P one.

In  order to account for the experimental observation of the shot noise in OMTJs, we have proposed \cite{cascales14}
a theoretical model based on tunneling through a two-level systems, like that presented above. Taking into account
the fact that the super-Poissonian shot noise appears mainly at larger voltages, such a two-level system may be
attributed to interfacial states of the PTCDA molecules in a biased junction.
Indeed, the experimental results can be quite well  explained qualitatively and also quantitatively in terms of
the model based on spin-dependent
electron tunneling through an interacting two-level system, described in detail in the preceding sections.
In order to qualitatively account for the experimentally observed situation with the Fano factor in the AP
state (on the average, we observe $F=1.5$) being  smaller than the Fano factor in the P state (1.7), we did
numerical  calculations based on the model presented above, see Fig.~2, and from fitting to the experimental
data we evaluated the parameters that reproduce the Fano factors in both configurations.

\section{Discussions}

The results of our calculations are in qualitative agreement with the physical interpretation
given in Ref.~\onlinecite{belzig05}. Indeed, considering the simplest two-level model (Fig.~1)
it was concluded that the
generating function $S(\chi )$ can be presented as a sum of independent Poissonian processes
of transferring $ne$ charges with probability of $(1/2)^{n}$ with $n=1$ to $\infty $.
In turn, the process of transferring $ne$ charges with large $n$ during one cycle
is possible because the tunneling to the left lead from the lower level
is strongly suppressed by the temperature factor $(1-f(\varepsilon _-))$.
In other words, several electrons can be quickly transferred through the
upper level till the cycle is stopped by an electron at the lower level.
This is a super-Poissonian process, and the Fano factor is equal to 3.

In our calculations we used the model of QDs with two energy levels, when
one of them is located below the Fermi level of left electrode $\varepsilon _{FL}$, and the other one
is between the Fermi levels of left and right electrodes. In reality the QD or
molecule can have many energy levels, with part of them situated below $\varepsilon _{FL}$
and another part between $\varepsilon _{FL}$ and $\varepsilon _{FR}$. It is rather
obvious that this is not so important for the mechanism of super-Poissonian
noise related to blocking of electron transport through the low energy level.
Generalization to the multilevel system with $N_+$ upper and $N_-$ lower levels
(bunched in two blocks with the same tunneling probability in each block) can
change the statistics, so that $F=(1+p)/(1-p)$, where $p=N_+/(N_++N_-)$. In
particular, for $p=1/2$ we obtain again $F=3$.
In this multilevel model, one can also get $F=2$ with $p=1/3$, which corresponds
to $N_-=2N_+$ (e.g., lower level
is twice degenerate and the upper one is nondegenerate).
In the case of nonmagnetic system, each of the levels is spin degenerate.
Thus, assuming equal tunneling probabilities for the spin-up and spin-down electrons,
one would get $p=1/2$ and $F=3$.

We also assumed that the tunneling probabilities are different for the lower
and upper levels. This changes essentially the result for the Fano factor because
the probability of transferring $ne$ electrons includes now the weight factor
of the ratio $(\Gamma ^-_R/\Gamma _R)^n$ since the probability of tunneling
of a single electron from the right lead to the upper level is not equal to $1/2$
anymore. In other words, the transfer of electrons through the upper level can be
not so quick due to a lower probability of the corresponding tunneling, and this
partially suppresses the super-Poissonian process as a sum of Poissonian processes with
the transfer of multiple charges.

Within this approach  one can also consider electron tunneling through a chain of
molecules in relatively thick junctions. Now the energy levels of different molecules are
not exactly at the same energy. First, because there is a potential slope within the junction,
which shifts correspondingly all the energy levels in the junction.
Second, due to inevitable disorder,
there exist some fluctuations of potential. This means that the intermolecular tunneling
can be possible only due to emission or absorption of appropriate phonons.
In this situation one can expect
that there is only one 'optimal' path of the electron transfer through the chain of
molecules, which uses a chosen number of the energy levels.
The probability of charge transfer through other pathes is exponentially small since
it requires substantial energy change at each intermolecular tunneling.
Hence, we come back to a Poissonian process of the transfer of a single charge
through the molecular chain. In this case we naturally obtain $F=1$.

It is also worth noting, that the super-Poissonian noise can appear due to other physical mechanisms as well, 
for example due to electron-phonon or electron-electron interactions.\cite{blanter00} However, the mechanism 
proposed by Belzig~\cite{belzig05} and  based on tunneling through two or more discrete levels  is the most 
appropriate one in our case. Indeed, the assumption of tunneling through discrete levels (with one low-energy level) 
is physically reasonable and justified. Moreover, this model explains the possibility of a rather strong 
enhancement of the Fano factor, and is also able to account for the experimental observations in the studied system.

\begin{acknowledgments}
This work was supported by the National Science Center in Poland as a research project
No.~DEC-2012/06/M/ST3/00042. We also gratefully acknowledge support by UAM-Santander collaborative project (2015/ASIA/04)
as well as by the Spanish MINECO (MAT2012-32743 and MAT2015-66000-P) grants and the Comunidad de Madrid through
NANOFRONTMAG-CM (S2013/MIT-2850). J.P.C. acknowledges support from the Fundacion Seneca (Region de Murcia)
posdoctoral fellowship (19791/PD/15)
\end{acknowledgments}

\appendix

\section{Calculation of the shot noise}

Using the expression for $\tilde{\lambda }(\chi )$
\begin{eqnarray}
\label{a1}
\tilde{\lambda }(\chi )=-b\pm \sqrt{b^2-c}
\end{eqnarray}
we find
\begin{eqnarray}
\label{a2}
&&\tilde{\lambda }'=-b'\pm \frac{2bb'-c'}{2\sqrt{b^2-c}}\, ,
\\
&&\tilde{\lambda }''=-b''\pm \frac{2(b')^2+2bb''-c''}{2\sqrt{b^2-c}}
\mp \frac{(2bb'-c')^2}{4(b^2-c)^{3/2}}\, .\hskip0.3cm
\end{eqnarray}
In the limit of $\chi \to 0$ we get
\begin{eqnarray}
\label{a4}
&&b\simeq \frac{\Gamma _{L\uparrow}\Gamma _{R\downarrow}^-+\Gamma _{L\downarrow}\Gamma _{R\uparrow}^-}
{2(\Gamma _{R\uparrow}^-+\Gamma _{R\downarrow}^-)},
\\
&&c\simeq 0,
\\
&&b'\simeq -\frac{i(\Gamma _{L\uparrow}\Gamma _{R\uparrow}^-+\Gamma _{L\downarrow}\Gamma _{R\downarrow}^-)}
{2(\Gamma _{R\uparrow}^-+\Gamma _{R\downarrow}^-)^2}
\nonumber \\
&&\hskip0.5cm
\times (\Gamma _{R\uparrow}+\Gamma _{R\downarrow}+\Gamma _{R\uparrow}^-+\Gamma _{R\downarrow}^-),
\\
&&b''=\frac{(\Gamma _{L\uparrow }^-\Gamma _{R\uparrow }^-+\Gamma _{L\downarrow }^-\Gamma _{R\downarrow }^-)
(\Gamma _{R\uparrow }^-+\Gamma _{R\downarrow }^-+\Gamma _{R\uparrow }^++\Gamma _{R\downarrow }^+)}
{(\Gamma _{R\uparrow }^-+\Gamma _{R\downarrow }^-)^2}
\nonumber \\
&&\hskip0.5cm
\times \left( \frac12 +\frac{\Gamma _{R\uparrow }^++\Gamma _{R\downarrow }^+}
{\Gamma _{R\uparrow }^-+\Gamma _{R\downarrow }^-}\right) ,
\\
&&c'\simeq -\frac{i\Gamma _{L\uparrow}\Gamma _{L\downarrow}
(\Gamma _{R\uparrow}+\Gamma _{R\downarrow}+\Gamma _{R\uparrow}^-+\Gamma _{R\downarrow}^-)}
{\Gamma _{R\uparrow}^-+\Gamma _{R\downarrow}^-},
\\
&&c''\simeq \frac{\Gamma _{L\uparrow}\Gamma _{L\downarrow}
(\Gamma _{R\uparrow}+\Gamma _{R\downarrow}+\Gamma _{R\uparrow}^-+\Gamma _{R\downarrow}^-)}
{(\Gamma _{R\uparrow}^-+\Gamma _{R\downarrow}^-)^2}
\nonumber \\
&&\hskip1cm \times
[2(\Gamma _{R\uparrow}+\Gamma _{R\downarrow})+\Gamma _{R\uparrow}^-+\Gamma _{R\downarrow}^-] .
\end{eqnarray}
Then we obtain the cumulants
\begin{eqnarray}
\label{a10}
C_1=iS'(\chi )_{\chi =0}
=\frac{x\Gamma _{L\uparrow}\Gamma _{L\downarrow}
(\Gamma _{R\uparrow}+\Gamma _{R\downarrow}+\Gamma _{R\uparrow}^-+\Gamma _{R\downarrow}^-)}
{\Gamma _{L\uparrow}\Gamma _{R\downarrow}^-+\Gamma _{L\downarrow}\Gamma _{R\uparrow}^-}\, ,\hskip0.6cm
\end{eqnarray}
\begin{eqnarray}
\label{a11}
C_2=S''(\chi )_{\chi =0}
=x\Gamma _{L\uparrow }\Gamma _{L\downarrow }
(\Gamma _{R\uparrow }+\Gamma _{R\uparrow }+\Gamma _{R\uparrow }^-+\Gamma _{R\downarrow }^-)
\nonumber \\ \times
\left[
\frac{2(\Gamma _{R\uparrow }+\Gamma _{R\downarrow })+\Gamma _{R\uparrow }^-+\Gamma _{R\downarrow }^-}
{(\Gamma _{R\uparrow }^-+\Gamma _{R\downarrow }^-)
(\Gamma _{L\uparrow }\Gamma _{R\downarrow }^-+\Gamma _{L\downarrow }\Gamma _{R\uparrow }^-)}
\right. \nonumber \\ \left.
+\frac{2(\Gamma _{L\uparrow }\Gamma _{R\uparrow }^-+\Gamma _{L\downarrow }\Gamma _{R\downarrow }^-)
(\Gamma _{R\uparrow }+\Gamma _{R\uparrow }+\Gamma _{R\uparrow }^-+\Gamma _{R\downarrow }^-)}
{(\Gamma _{R\uparrow }^-+\Gamma _{R\downarrow }^-)
(\Gamma _{L\uparrow }\Gamma _{R\downarrow }^-+\Gamma _{L\downarrow }\Gamma _{R\uparrow }^-)^2}
\right. \nonumber \\ \left.
-\frac{2\Gamma _{L\uparrow }\Gamma _{L\downarrow }
(\Gamma _{R\uparrow }+\Gamma _{R\uparrow }+\Gamma _{R\uparrow }^-+\Gamma _{R\downarrow }^-)
(\Gamma _{R\uparrow }^-+\Gamma _{R\downarrow }^-)}
{(\Gamma _{L\uparrow }\Gamma _{R\downarrow }^-+\Gamma _{L\downarrow }\Gamma _{R\uparrow }^-)^3}
\right] \hskip0.5cm
\end{eqnarray}
and the explicit formula for shot noise $S_2$
\begin{eqnarray}
\label{a12}
S_2=2e^2x\Gamma ^-_{L\uparrow }\Gamma ^-_{L\downarrow }
\big( \Gamma ^+_{R\uparrow }+\Gamma ^+_{R\downarrow }+\Gamma _{R\uparrow }^-+\Gamma _{R\downarrow }^-\big)
\hskip1.5cm
\nonumber \\ \times
\left[
\frac{2(\Gamma ^+_{R\uparrow }+\Gamma ^+_{R\downarrow })+\Gamma _{R\uparrow }^-+\Gamma _{R\downarrow }^-}
{(\Gamma _{R\uparrow }^-+\Gamma _{R\downarrow }^-)
(\Gamma _{L\uparrow }^-\Gamma _{R\downarrow }^-+\Gamma _{L\downarrow }^-\Gamma _{R\uparrow }^-)}
\right. \nonumber \\ \left.
+\frac{2(\Gamma ^-_{L\uparrow }\Gamma _{R\uparrow }^-+\Gamma ^-_{L\downarrow }\Gamma _{R\downarrow }^-)
(\Gamma ^+_{R\uparrow }+\Gamma ^+_{R\downarrow }+\Gamma _{R\uparrow }^-+\Gamma _{R\downarrow }^-)}
{(\Gamma _{R\uparrow }^-+\Gamma _{R\downarrow }^-)
(\Gamma ^-_{L\uparrow }\Gamma _{R\downarrow }^-+\Gamma ^-_{L\downarrow }\Gamma _{R\uparrow }^-)^2}
\right. \nonumber \\ \left.
-\frac{2\Gamma ^-_{L\uparrow }\Gamma ^-_{L\downarrow }
(\Gamma ^+_{R\uparrow }+\Gamma ^+_{R\downarrow }+\Gamma _{R\uparrow }^-+\Gamma _{R\downarrow }^-)
(\Gamma _{R\uparrow }^-+\Gamma _{R\downarrow }^-)}
{(\Gamma ^-_{L\uparrow }\Gamma _{R\downarrow }^-+\Gamma ^-_{L\downarrow }\Gamma _{R\uparrow }^-)^3}
\right] .\hskip0.6cm
\end{eqnarray}


\begin{thebibliography}{99}
\bibitem{molecule}
We call it quantum dot but it can be also a molecule if we do not
account for possible phonon modes coupled to electrons states in molecules.
\end{thebibliography}
\end{document}